\documentclass[cameraready]{Interspeech}
\usepackage{graphicx}
\usepackage{float}
\usepackage{subcaption}
\usepackage[dvipsnames]{xcolor}
\title{Factors affecting ASR performance: A study using state of the art ASR models in Indic Languages}

\author[affiliation={1}]{Agneedh}{Basu}
\author[affiliation={1}]{Pavan K}{J}
\author[affiliation={1}]{Pranav}{Bhat}
\author[affiliation={1}]{Sujith}{P}
\author[affiliation={1}]{Visruth}{Sanka}
\author[affiliation={1}]{Nihar}{Desai}
\author[affiliation={2}]{Prasanta K}{Ghosh}
\address{
    $^1$ AI \& Robotics Technology Park (ARTPARK), I-Hub @ IISc, Bangalore, India \\
    $^2$Department of Electrical Engineering, Indian Institute of Science, Bangalore, India
}

\email{agneedh@artpark.in, pavanjk@artpark.in, sujith@artpark.in}


\usepackage{comment}

\begin{document}

\maketitle

\begin{abstract}
ASR performance varies across languages, speakers, and recording conditions, yet systematic analysis for Indic languages remain limited. We present a large-scale study of decoded outputs from multiple open-source ASR models evaluated on diverse Indian speech datasets in zero-shot settings. We analyze linguistic, speaker-level, and acoustic factors across Hindi, Bengali, Kannada, Telugu, and Marathi. We examine correlations between WER and speaker traits such as average word length, speaking rate, and utterance duration across multiple model–dataset pairs. For Hindi, we further analyze audio factors including telephone codecs, bit depth, resampling, and background noise. Results reveal both cross-lingual patterns and language-specific sensitivities, showing how speaker behavior and signal processing choices affect ASR robustness in real-world Indic scenarios.
\end{abstract}

\section{Introduction}
Automatic Speech Recognition (ASR) has advanced significantly due to deep learning methods such as Connectionist Temporal Classification (CTC) \cite{graves2006ctc}, attention-based encoder–decoder models \cite{graves2013speech}, and Transformer architectures \cite{vaswani2017attention}. Despite these advances, robust performance across linguistically diverse and under-resourced languages remains challenging \cite{pratap2020massively}. Indic languages — spoken by over a billion people — pose particular difficulties due to rich phonetic inventories, complex morphology, dialectal diversity, and variability in recording conditions \cite{vakyansh}.
While ASR systems achieve high accuracy for languages such as English and Mandarin Chinese, Indic languages including Hindi, Bengali, Kannada, Telugu, and Marathi often lack standardized large-scale corpora and comprehensive evaluation benchmarks \cite{pratap2020massively}. Differences in speaker demographics, speaking styles, and acoustic environments further increase performance variability in real-world settings \cite{li2014noise}. Prior work on Indic ASR has largely focused on corpus development and baseline modeling \cite{jha2018indic}, language-specific improvements through phoneme hybrid systems \cite{rao2016acoustic}, and multilingual transfer learning \cite{singh2021multilingual}, with some exploration of noise and channel robustness \cite{banerjee2017noise}. However, these studies examine factors in isolation and for limited language sets, leaving a unified cross-lingual analysis largely absent.
While individual factors such as speaking rate, word length, and bandwidth degradation are known to affect ASR, a joint multi-factor, multi-model analysis across typologically diverse Indic languages has not been undertaken. To the best of our knowledge, this is the first study to systematically examine both speaker-level and audio-level factors across five Indic languages and multiple state-of-the-art ASR architectures in zero-shot settings. We analyze correlations between WER and speaker traits including average word length, speaking rate, and utterance duration. For Hindi — the most widely spoken and a representative high-resource Indic language — we further investigate signal-level factors such as telephone codecs, amplitude precision, resampling strategies, and additive noise. Through this unified analysis, we provide a comprehensive empirical baseline for understanding and improving ASR robustness in real-world Indic deployment scenarios.

\section{Experimental Setup}

\subsection{Experimental Factors}
We analyze both speaker-level and audio-level factors that are hypothesized to influence ASR performance. All analyses are conducted using Word Error Rate (WER) as the primary evaluation metric.

\subsubsection{Speaker-Level Factors}

Speaker-level factors are derived from reference transcriptions and capture linguistic and speaking-style characteristics. We consider three features: (1) Average Word Length (AWL), defined as the mean number of characters per word, serving as a proxy for morphological complexity across Indic languages; and (2) Speaking Rate (Words Per Minute, WPM), computed as the number of words divided by utterance duration. Variations in speaking rate can affect acoustic modeling and alignment quality, particularly in spontaneous speech; (3) Utterance Duration (Audio Length, AL), to analyze WER trends across short and long utterances. These features provide interpretable proxies for linguistic complexity and temporal variability

\subsubsection{Audio-Level Factors}
For Hindi, we further evaluate ASR robustness under diverse signal conditions. Starting from 16 kHz, 16-bit mono audio, we apply controlled degradations at inference time only to simulate realistic telephony and low-fidelity recording scenarios. The transformation factors are:

\textbf{Telephone Codecs:} We simulate cellular transmission conditions using standard speech codecs. GSM (2G) is modeled by downsampling to 8 kHz, performing GSM encode–decode, and resampling to 16 kHz to reflect severe bandwidth constraints \cite{etsi_gsm610}. Narrowband (3G) speech is simulated via 3.4 kHz low-pass filtering and 8 kHz resampling to isolate bandwidth limitation without codec artifacts \cite{itu_g712}. Wideband (4G) speech is modeled using 8 kHz low-pass filtering and 16 kHz resampling to preserve partial high-frequency content \cite{itu_g722}. Opus (5G) transmission is simulated through Opus encode–decode processing followed by resampling to 16 kHz \cite{valin2012opus}.

\textbf{Upsampling Methods:} Audio is first downsampled to 4 kHz and 8 kHz and then restored to 16 kHz using different techniques. Classical baselines include high-quality sinc-based polyphase resampling (soxr\_hq) \cite{smith2002digital} and linear interpolation \cite{oppenheim1999discrete}. We also evaluate neural restoration approaches: VoiceFixer, which performs bandwidth extension in the time–frequency domain \cite{liu2022voicefixer}, and AudioSR, a diffusion-based super-resolution model designed to reconstruct perceptually realistic high-frequency components \cite{liu2023audiosr}.

\textbf{Bit Precision:} To assess sensitivity to amplitude resolution, 16-bit PCM audio is uniformly quantized to 12, 10, 8, and 6 bits. This introduces controlled quantization noise while preserving sampling rate and temporal structure.

\textbf{Additive Noise:} Robustness is evaluated under three noise conditions: white Gaussian noise for controlled SNR analysis; background audio created by mixing competing human speech to simulate overlapping talkers; and natural environmental non-speech sounds sourced from the AudioSet-NonSpeech dataset\footnote{\url{https://huggingface.co/datasets/bond005/audioset-nonspeech}}, reflecting realistic deployment environments.

\subsection{ASR Models}
We evaluate recent open-source ASR systems based on Whisper, Wav2Vec2, data2vec, and Conformer architectures, following a self-supervised pre-training and fine-tuning paradigm. Our study includes both Indic-specialized and multilingual models. The models used are: Indic Conformer \cite{indic_conformer}, data2vec-aqc \cite{data2cev-aqc, baevski2022data2vec}, Vakyansh Toolkit \cite{vakyansh}, Vaani Whisper \cite{vaani-whisper}, Voxtral Mini \cite{liu2025voxtral}, Shrutam-HindiASR-1.0 \cite{bharatgenai2025shrutam}, and OpenAI Whisper-large-v3 \cite{whisper}. For the audio-level robustness experiments, we use Indic Conformer's Hindi model (Indic-Conformer-hi) and Vaani's whisper-large-v3 for Hindi model (Vaani-Whisper-L-hi).

\subsection{Evaluation Datasets}
To capture broad linguistic variation and recording circumstances, we evaluate across the test splits of the following datasets: MUCS \cite{mucs}, Kathbath \cite{kathbath}, IndicTTS \cite{IITM_IndicTTS}, Common Voice \cite{commonvoice}, FLEURS \cite{fleurs2022arxiv}, Vaani \cite{vaani2025}, and RESPIN \cite{saurabh2025respin}. For audio-level factor analysis, we use the Hindi test sets of FLEURS and Kathbath.

\section{Results and Observations}
We evaluate the aforementioned open-source ASR models on the described Indic speech datasets using word error rate (WER) as the primary metric. Prior to scoring, both hypotheses and references are normalized by removing punctuation, tags, special tokens, and other non-verbal artifacts.

\subsection{Effect of Speaker-Level Factors}

We examine the impact of average word length (AWL), speaking rate (WPM), and utterance duration (AL) on ASR performance. For each factor–language pair, WER trends are plotted across all model–dataset combinations (light lines), with a darker line indicating the averaged trend, revealing both model variability and consistent factor-driven effects.

\begin{figure*}[!t]
    \centering

    % Top row (3 plots)
    \begin{subfigure}[b]{0.32\textwidth}
        \centering
        \includegraphics[width=\textwidth]{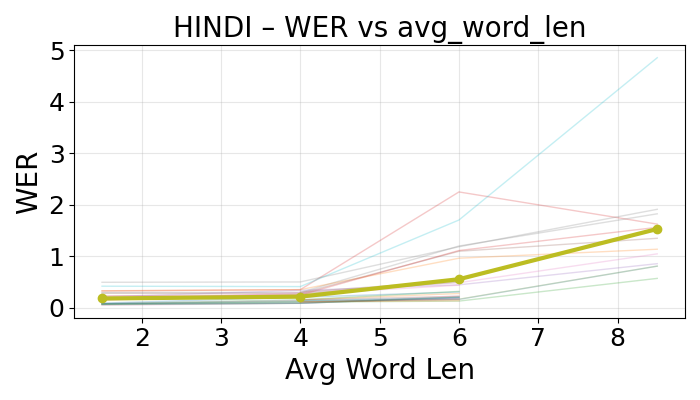}
        \caption{Hindi - AWL}
    \end{subfigure}
    \hfill
    \begin{subfigure}[b]{0.32\textwidth}
        \centering
        \includegraphics[width=\textwidth]{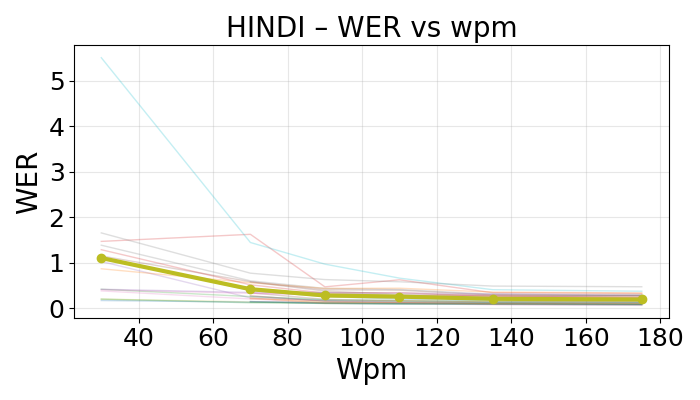}
        \caption{Hindi - WPM}
    \end{subfigure}
    \hfill
    \begin{subfigure}[b]{0.32\textwidth}
        \centering
        \includegraphics[width=\textwidth]{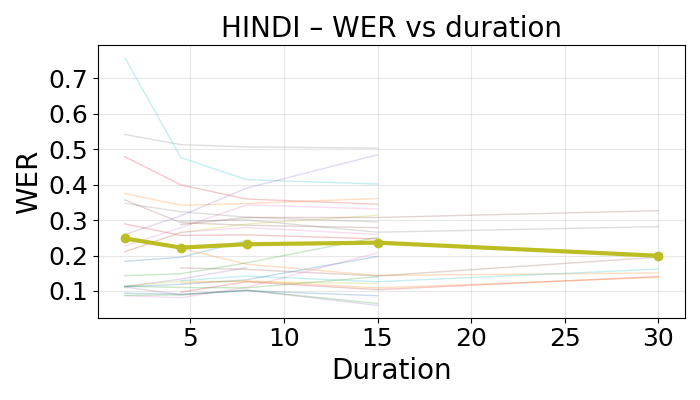}
        \caption{Hindi - AL}
    \end{subfigure}

    \vspace{0.3em}

    % 2nd row (3 plots)
    \begin{subfigure}[b]{0.32\textwidth}
        \centering
        \includegraphics[width=\textwidth]{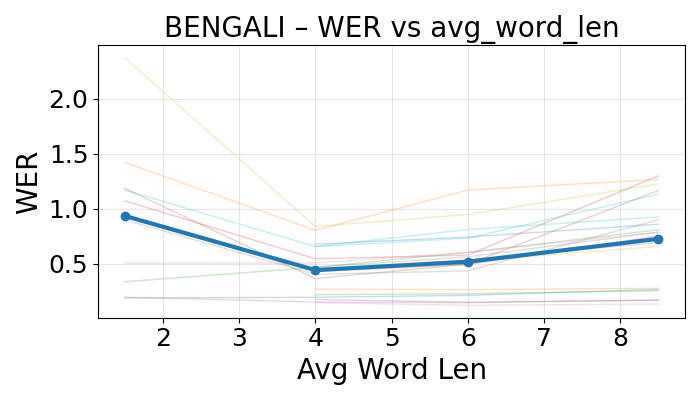}
        \caption{Bengali - AWL}
    \end{subfigure}
    \hfill
    \begin{subfigure}[b]{0.32\textwidth}
        \centering
        \includegraphics[width=\textwidth]{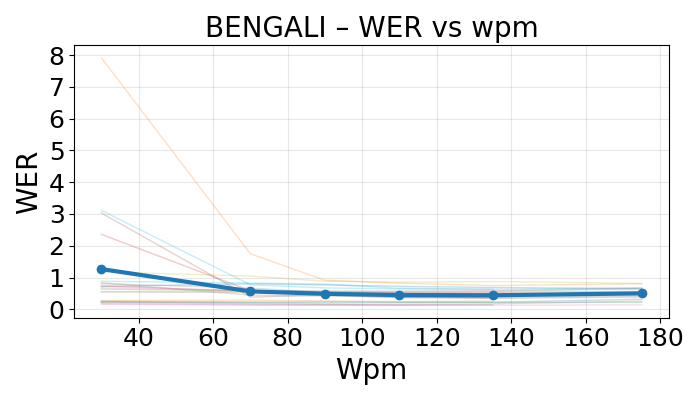}
        \caption{Bengali - WPM}
    \end{subfigure}
    \hfill
    \begin{subfigure}[b]{0.32\textwidth}
        \centering
        \includegraphics[width=\textwidth]{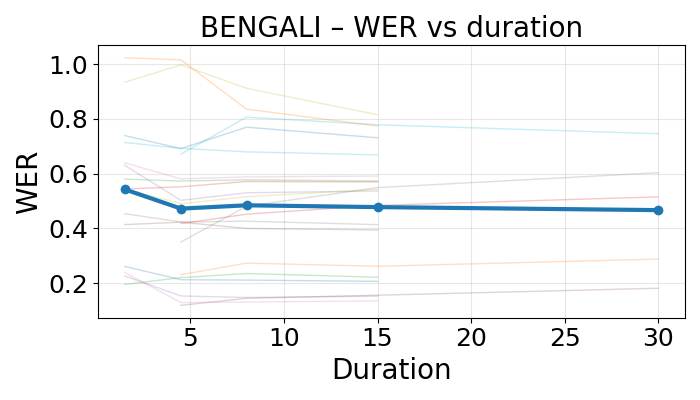}
        \caption{Bengali - AL}
    \end{subfigure}

    % 3rd row (3 plots)
    \begin{subfigure}[b]{0.32\textwidth}
        \centering
        \includegraphics[width=\textwidth]{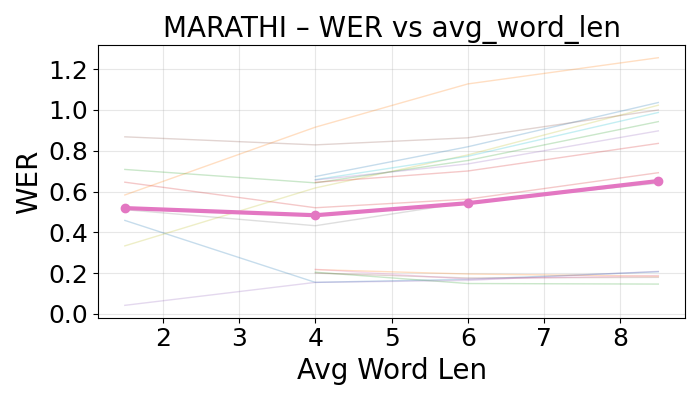}
        \caption{Marathi - AWL}
    \end{subfigure}
    \hfill
    \begin{subfigure}[b]{0.32\textwidth}
        \centering
        \includegraphics[width=\textwidth]{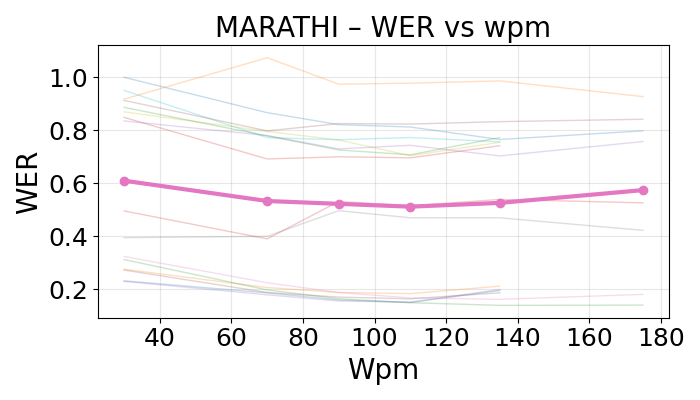}
        \caption{Marathi - WPM}
    \end{subfigure}
    \hfill
    \begin{subfigure}[b]{0.32\textwidth}
        \centering
        \includegraphics[width=\textwidth]{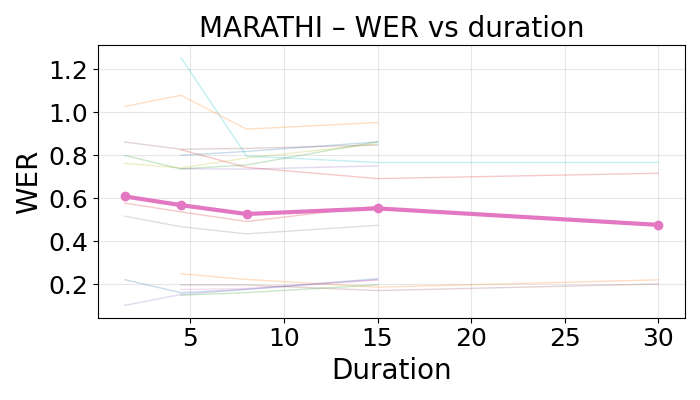}
        \caption{Marathi - AL}
    \end{subfigure}

    \begin{subfigure}[b]{0.32\textwidth}
        \centering
        \includegraphics[width=\textwidth]{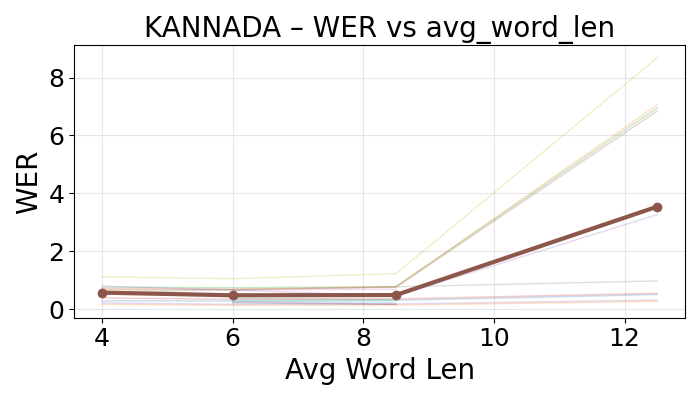}
        \caption{Kannada - AWL}
    \end{subfigure}
    \hfill
    \begin{subfigure}[b]{0.32\textwidth}
        \centering
        \includegraphics[width=\textwidth]{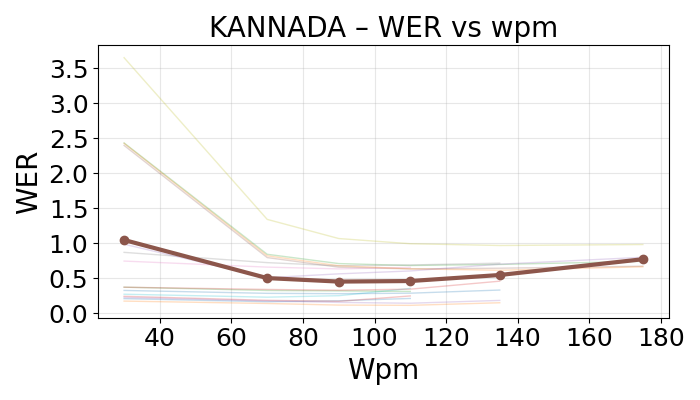}
        \caption{Kannada - WPM}
    \end{subfigure}
    \hfill
    \begin{subfigure}[b]{0.32\textwidth}
        \centering
        \includegraphics[width=\textwidth]{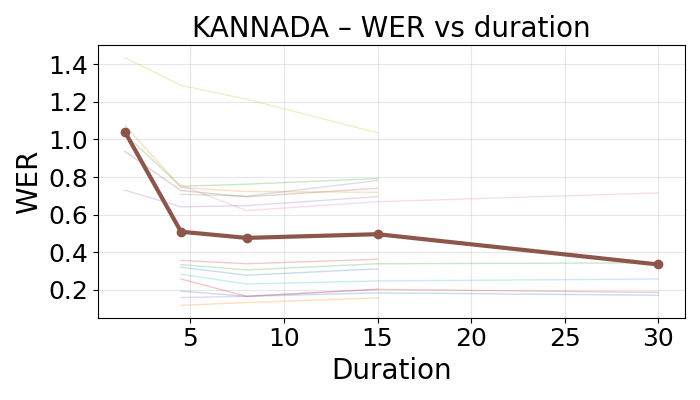}
        \caption{Kannada - AL}
    \end{subfigure}

    \begin{subfigure}[b]{0.32\textwidth}
        \centering
        \includegraphics[width=\textwidth]{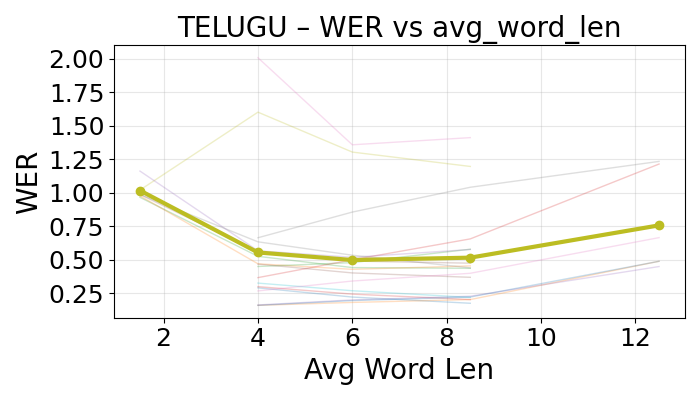}
        \caption{Telugu - AWL}
    \end{subfigure}
    \hfill
    \begin{subfigure}[b]{0.32\textwidth}
        \centering
        \includegraphics[width=\textwidth]{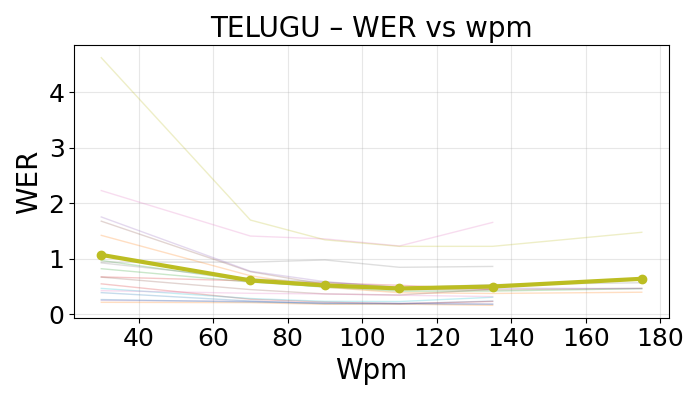}
        \caption{Telugu - WPM}
    \end{subfigure}
    \hfill
    \begin{subfigure}[b]{0.32\textwidth}
        \centering
        \includegraphics[width=\textwidth]{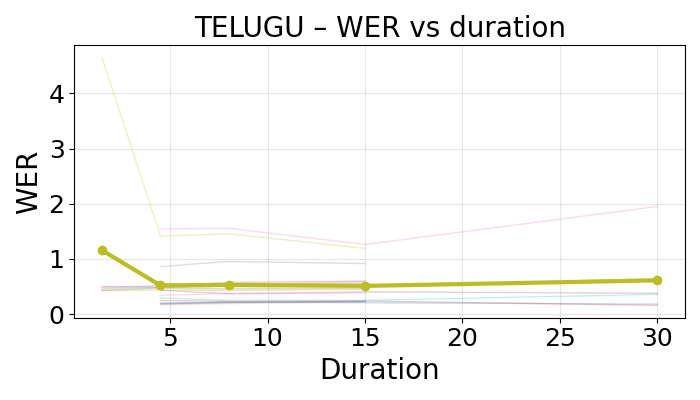}
        \caption{Telugu - AL}
    \end{subfigure}

    \caption{
    Impact of speaker-level factors across languages. Light-colored lines denote WER trends for individual model--dataset combinations, while the dark line represents the average trend across all models and datasets.
    }
    \label{fig:avg_wordlen_trends}
\end{figure*}

\begin{enumerate}
    \item \textbf{Average Word Length (Figure~\ref{fig:avg_wordlen_trends}):} WER generally decreases for short-to-medium word lengths and increases for longer words. The consistency of this pattern across models suggests that word length is an intrinsic source of ASR errors rather than a model-specific artifact.

    \item \textbf{Speaking Rate (Figure~\ref{fig:avg_wordlen_trends}):} The relationship between WPM and WER is non-monotonic and language-dependent. In Hindi, WER decreases with increasing speaking rate, with slow speech showing higher errors, possibly due to hesitations or noise. Other languages exhibit degradation at higher speaking rates, indicating sensitivity to rapid articulation.

    \item \textbf{Utterance Duration (Figure~\ref{fig:avg_wordlen_trends}):} WER increases for very short utterances due to limited phonetic context and boundary effects, while longer utterances show gradual degradation from error accumulation. These trends remain consistent across architectures.
\end{enumerate}

%%%%%%%%%%%%%%%%%%%%%%%%%%%%%%%%%%%%%%%%%%%%%%%%%

\subsection{Effect of Audio-Level Factors}

We analyze audio-level robustness by plotting WER across model–dataset combinations under controlled degradations.

\begin{figure*}[!t]
    \centering
    \begin{subfigure}[b]{0.48\textwidth}
        \centering
        \includegraphics[width=\textwidth]{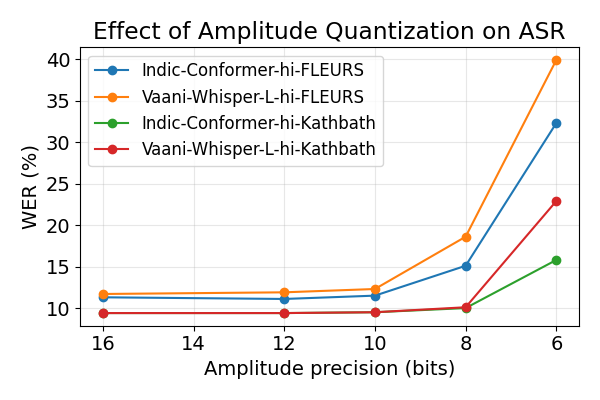}
        \caption{Effect of amplitude quantization on ASR performance.}
        \label{fig:quantization}
    \end{subfigure}
    \hfill
    \begin{subfigure}[b]{0.48\textwidth}
        \centering
        \includegraphics[width=\textwidth]{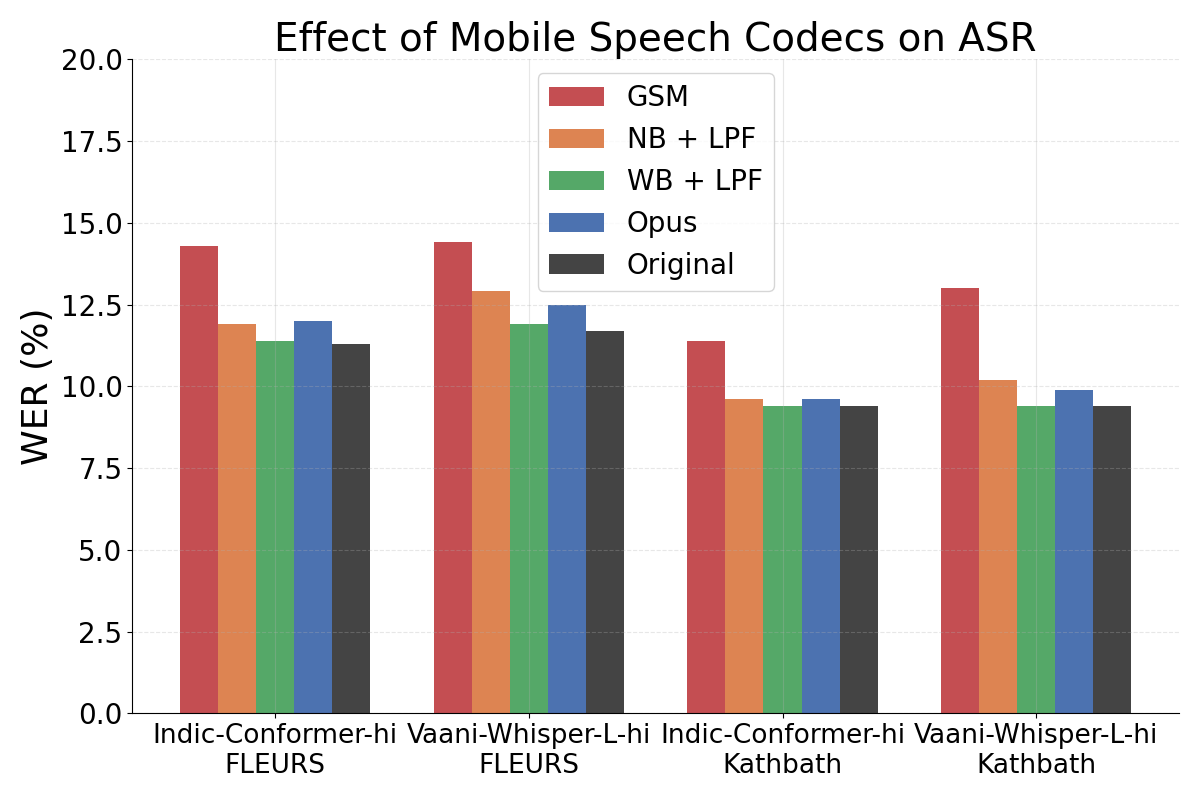}
        \caption{Effect of mobile speech codecs on ASR performance.}
        \label{fig:codecs}
    \end{subfigure}
    \caption{Impact of signal precision and communication codecs on ASR WER across models and datasets.}
    \label{fig:quant_codec}
\end{figure*}

\begin{figure*}[!t]
    \centering
    \begin{subfigure}[b]{0.48\textwidth}
        \centering
        \includegraphics[width=\textwidth]{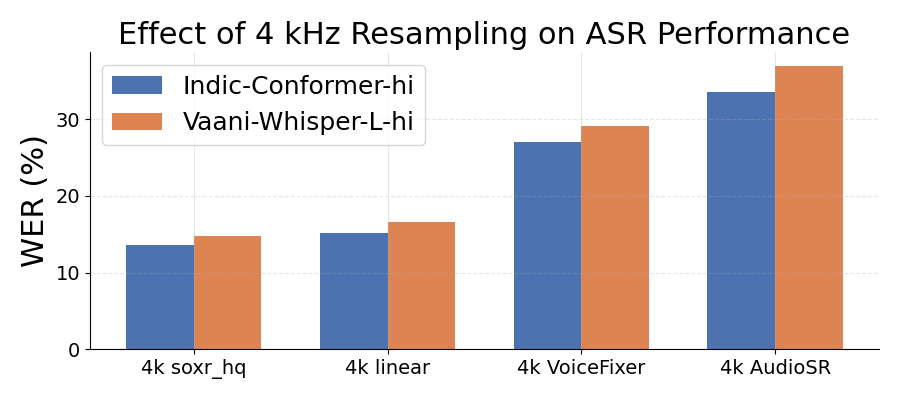}
        \caption{4 kHz downsampling and restoration methods.}
        \label{fig:4k_resample}
    \end{subfigure}
    \hfill
    \begin{subfigure}[b]{0.48\textwidth}
        \centering
        \includegraphics[width=\textwidth]{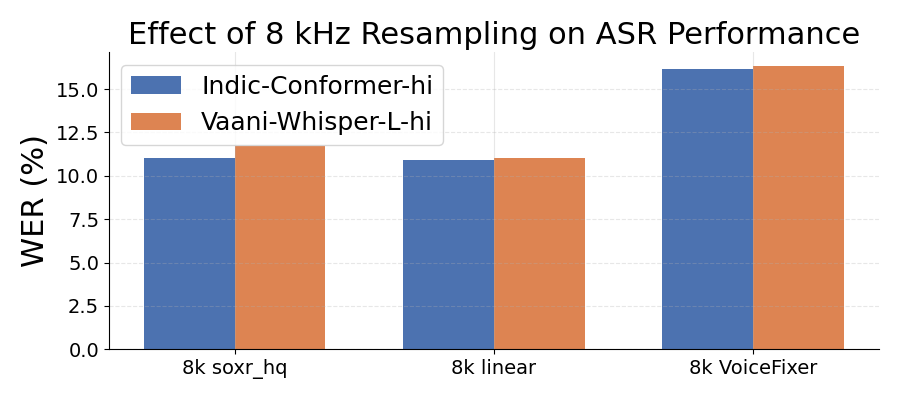}
        \caption{8 kHz downsampling and restoration methods.}
        \label{fig:8k_resample}
    \end{subfigure}
    \caption{Effect of aggressive (4 kHz) and moderate (8 kHz) resampling and different reconstruction techniques on ASR performance. WER values are mean across Kathbath and FLEURS datasets.}
    \label{fig:resampling}
\end{figure*}

\begin{figure*}[!t]
\centering

% Row of three plots
\begin{subfigure}{0.32\textwidth}
    \centering
    \includegraphics[width=\linewidth]{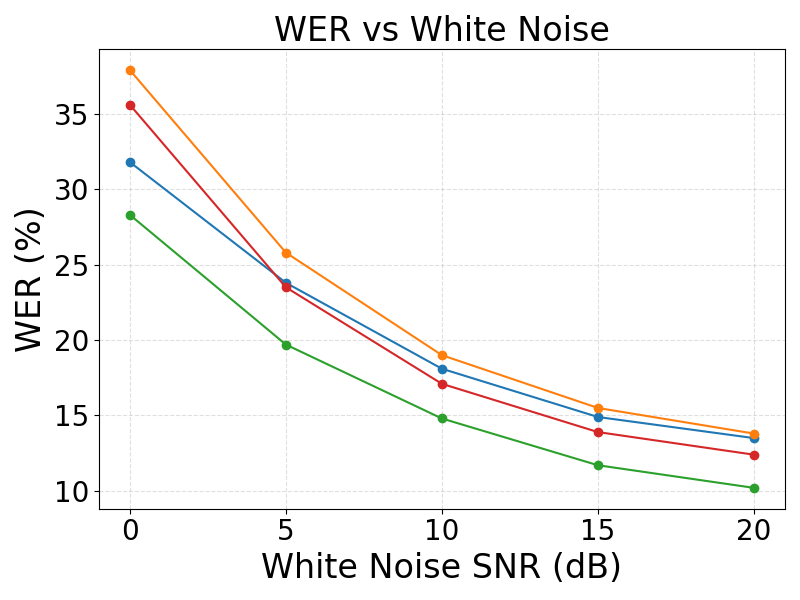}
    \caption{White Noise}
\end{subfigure}
\hfill
\begin{subfigure}{0.32\textwidth}
    \centering
    \includegraphics[width=\linewidth]{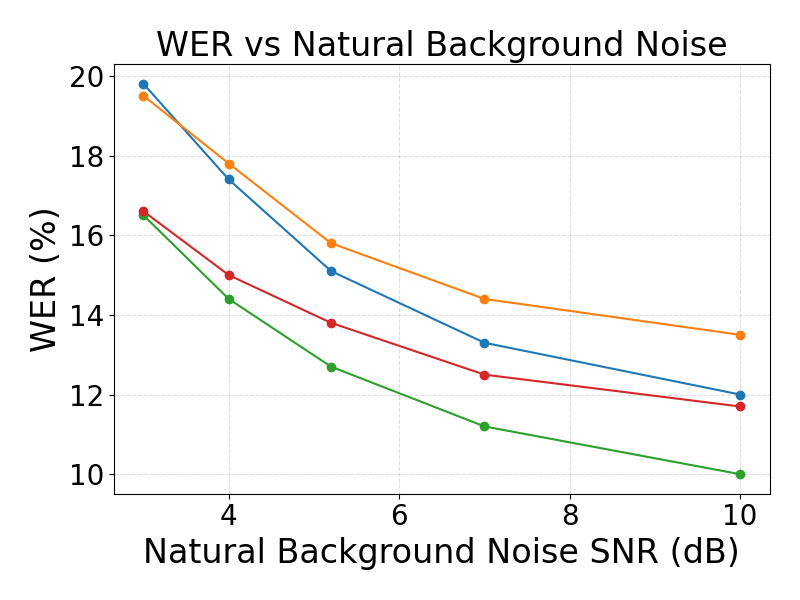}
    \caption{Natural Background Noise}
\end{subfigure}
\hfill
\begin{subfigure}{0.32\textwidth}
    \centering
    \includegraphics[width=\linewidth]{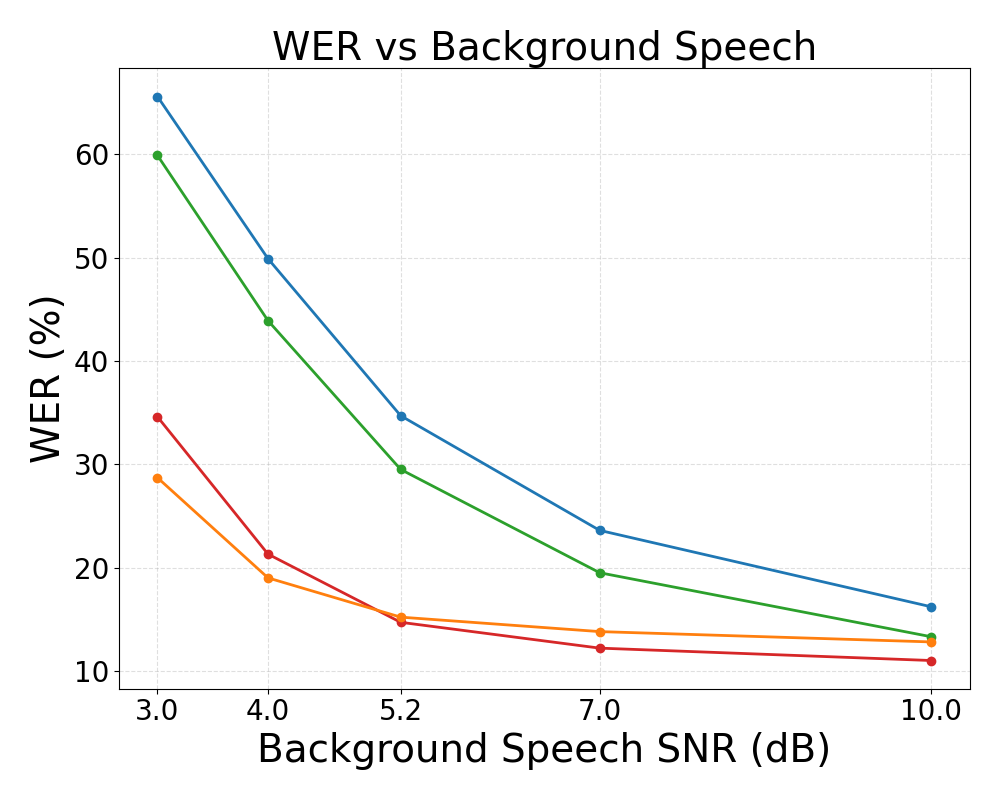}
    \caption{Background Speech}
\end{subfigure}

\vspace{0.2cm}

% Common Legend (manual)
\begin{center}
\footnotesize
\textbf{Models:} 
% \textcolor[HTML]{1f77b4}{\rule{0.8em}{0.8em}} Indic-Conformer-hi-FLEURS \quad
% \textcolor{orange}{\rule{0.8em}{0.8em}} Vaani-Whisper-L-hi-FLEURS \quad
% \textcolor{green}{\rule{0.8em}{0.8em}} Indic-Conformer-hi-Kathbath \quad
% \textcolor{red}{\rule{0.8em}{0.8em}} Vaani-Whisper-L-hi-Kathbath
\textcolor[HTML]{1f77b4}{\rule{0.8em}{0.8em}} Indic-Conformer-hi-FLEURS \quad
\textcolor[HTML]{ff7f0e}{\rule{0.8em}{0.8em}} Vaani-Whisper-L-hi-FLEURS \quad
\textcolor[HTML]{2ca02c}{\rule{0.8em}{0.8em}} Indic-Conformer-hi-Kathbath \quad
\textcolor[HTML]{d62728}{\rule{0.8em}{0.8em}} Vaani-Whisper-L-hi-Kathbath

\end{center}

\caption{Effect of additive interference on ASR performance for Hindi. 
(a) stationary white noise, (b) natural background environmental noise, (c) competing background speech. 
}
\label{fig:bg_noise}
\end{figure*}

\begin{enumerate}

\item \textbf{Amplitude Precision (Figure~\ref{fig:quantization}):} ASR systems remain stable under 10–12 bit quantization but degrade sharply at 8 bits and below. At 6 bits, WER increases substantially across all models, indicating a critical precision threshold. These results show robustness to moderate quantization but strong sensitivity to aggressive amplitude reduction that distorts spectral cues.

\item \textbf{Mobile Speech Codecs (Figure~\ref{fig:codecs}):} GSM (2G) consistently degrades performance due to narrowband constraints and quantization. Simulated Narrowband(3G) / Wideband(4G) conditions preserve accuracy close to the original 16 kHz audio, while Opus (5G) introduces only marginal degradation. 

\item \textbf{Upsampling Methods (Figure~\ref{fig:resampling}):} Classical resampling (linear, soxr hq) yields moderate loss, whereas neural restoration methods (VoiceFixer, AudioSR) paradoxically worsen WER despite producing perceptually superior audio. This is likely because these models are optimized for human auditory perception — reconstructing high-frequency components that sound natural — rather than preserving the spectral structure that acoustic models rely on for phoneme discrimination. The enhanced output may introduce hallucinated frequency content that acts as a confounding artifact during ASR decoding.

\item \textbf{Additive Noise (Figure~\ref{fig:bg_noise}):} For white Gaussian noise, background speech and natural environmental background noise, WER follows the expected trend, degrading at low SNRs and improving steadily as SNR increases, indicating reasonable robustness to stationary and non-speech noise. Under the background speech setting, Whisper-based models demonstrate greater robustness than Conformer-based models, suggesting improved handling of overlapping speech compared to purely acoustic distortions.

\end{enumerate}

%%%%%%%%%%%%%%%%%%%%%%%%%%%%%%%%%%%%%%%%%%%%%%%%%%%%%%

\section{Conclusion}
We present a unified analysis of speaker- and audio-level factors affecting Indic ASR under zero-shot settings, with direct implications for telephony and real-world deployment. Speaker-level trends show consistent WER degradation as conditions worsen, with similar trajectories across models and datasets, indicating systematic rather than speaker-specific effects.
Among audio factors, bandwidth limitation is the most critical: narrowband conditions (e.g., 2G GSM) significantly degrade performance, while wideband transmission (3G–5G) preserves accuracy close to clean baselines — indicating that telephony-grade ASR pipelines should prioritize wideband codec support. Quantization experiments reveal a precision threshold, with 10–12 bit audio remaining stable and sharp degradation at 8-bit and below, relevant for edge deployment constraints.
Neural restoration methods may introduce harmful artifacts, making classical resampling more reliable for telephony preprocessing. Competing background speech causes the strongest degradation, with Whisper models showing comparatively better robustness — a useful criterion for model selection in call-center environments.
Overall, preserving bandwidth and amplitude precision at the transmission stage is more impactful than perceptual enhancement, providing actionable guidance for robust Indic ASR deployment across diverse communication infrastructure.

\bibliographystyle{IEEEtran}
\bibliography{mybib}

@inproceedings{graves2013speech,
  title={Speech recognition with deep recurrent neural networks},
  author={Graves, Alex and Mohamed, Abdel-rahman and Hinton, Geoffrey},
  booktitle={Proc. ICASSP},
  pages={6645--6649},
  year={2013}
}

@inproceedings{vaswani2017attention,
  title={Attention is all you need},
  author={Vaswani, Ashish and Shazeer, Noam and Parmar, Niki and Uszkoreit, Jakob and Jones, Llion and Gomez, Aidan N. and Kaiser, {\L}ukasz and Polosukhin, Illia},
  booktitle={Advances in Neural Information Processing Systems (NeurIPS)},
  pages={5998--6008},
  year={2017}
}

@inproceedings{jha2018indic,
  title={Indic speech recognition: Challenges and approaches},
  author={Jha, Manish and Mittal, Aayush and Dey, Subhadeep},
  booktitle={Proc. SLTU},
  year={2018}
}

@inproceedings{rao2016acoustic,
  title={Acoustic modeling for Indian languages},
  author={Rao, K. Sreenivasa and Murthy, Hema A.},
  booktitle={Proc. INTERSPEECH},
  pages={2726--2730},
  year={2016}
}

@inproceedings{singh2021multilingual,
  title={Multilingual transfer learning for low-resource Indian languages},
  author={Singh, Aman and Sitaram, Sunayana and Black, Alan W.},
  booktitle={Proc. INTERSPEECH},
  pages={1917--1921},
  year={2021}
}

@inproceedings{banerjee2017noise,
  title={Noise and channel robust speech recognition for Indian languages},
  author={Banerjee, Suman and Ghosh, Anupam and Rao, K. Sreenivasa},
  booktitle={Proc. INTERSPEECH},
  pages={2731--2735},
  year={2017}
}

@article{vakyansh,
  author       = {Anirudh Gupta and
                  Harveen Singh Chadha and
                  Priyanshi Shah and
                  Neeraj Chimmwal and
                  Ankur Dhuriya and
                  Rishabh Gaur and
                  Vivek Raghavan},
  title        = {{CLSRIL-23:} Cross Lingual Speech Representations for Indic Languages},
  journal      = {CoRR},
  volume       = {abs/2107.07402},
  year         = {2021},
  url          = {https://arxiv.org/abs/2107.07402},
  eprinttype    = {arXiv},
  eprint       = {2107.07402},
  bibsource    = {dblp computer science bibliography, https://dblp.org}
}

@inproceedings{commonvoice,
  author = {Ardila, R. and Branson, M. and Davis, K. and Henretty, M. and Kohler, M. and Meyer, J. and Morais, R. and Saunders, L. and Tyers, F. M. and Weber, G.},
  title = {Common Voice: A Massively-Multilingual Speech Corpus},
  booktitle = {Proceedings of the 12th Conference on Language Resources and Evaluation (LREC 2020)},
  pages = {4211--4215},
  year = 2020
}

@article{fleurs2022arxiv,
  title = {FLEURS: Few-shot Learning Evaluation of Universal Representations of Speech},
  author = {Conneau, Alexis and Ma, Min and Khanuja, Simran and Zhang, Yu and Axelrod, Vera and Dalmia, Siddharth and Riesa, Jason and Rivera, Clara and Bapna, Ankur},
  journal={arXiv preprint arXiv:2205.12446},
  url = {https://arxiv.org/abs/2205.12446},
  year = {2022},
}

@misc{kathbath,
  doi = {10.48550/ARXIV.2208.11761},
  url = {https://arxiv.org/abs/2208.11761},
  author = {Javed, Tahir and Bhogale, Kaushal Santosh and Raman, Abhigyan and Kunchukuttan, Anoop and Kumar, Pratyush and Khapra, Mitesh M.},
  title = {IndicSUPERB: A Speech Processing Universal Performance Benchmark for Indian languages},
  publisher = {arXiv},
  year = {2022},
  copyright = {arXiv.org perpetual, non-exclusive license}
}

@article{liu2025voxtral,
  title={Voxtral},
  author={Liu, Alexander H and Ehrenberg, Andy and Lo, Andy and Denoix, Cl{\'e}ment and Barreau, Corentin and Lample, Guillaume and Delignon, Jean-Malo and Chandu, Khyathi Raghavi and von Platen, Patrick and Muddireddy, Pavankumar Reddy and others},
  journal={arXiv preprint arXiv:2507.13264},
  year={2025}
}

@misc{saurabh2025respin,
  author       = {{Saurabh Kumar et al}},
  title        = {{RESPIN Corpus}: A Read‑Speech Corpus of 10,000+ Hours in Dialects of Nine Indian Languages},
  year         = {2025},
  howpublished = {Available at \url{https://spiredatasets.ee.iisc.ac.in/respincorpus}},
  note         = {Open‑source corpus developed by IISc RESPIN project, covering read speech across 9 Indian languages}  
}

@inproceedings{whisper,
  author    = {Alec Radford and Jong Wook Kim and Tao Xu and Greg Brockman and Christine McLeavey and Ilya Sutskever},
  title     = {Robust Speech Recognition via Large-Scale Weak Supervision},
  booktitle = {Proceedings of the 40th International Conference on Machine Learning (ICML)},
  year      = {2023},
  volume    = {202},
  pages     = {28492--28518},
  publisher = {PMLR},
  url       = {https://proceedings.mlr.press/v202/radford23a.html}
}

@misc{bharatgenai2025shrutam,
  author = {BharatGenAI},
  title = {Shrutam-HindiASR-1.0: Hindi Automatic Speech Recognition Model},
  year = {2025},
  publisher = {Hugging Face},
  note = {Accessed: Sep. 29, 2025},
  url = {https://huggingface.co/bharatgenai/Shrutam-HindiASR-1.0}
}

@misc{data2cev-aqc,
  author = {Vasista Sai Lodagala and Sreyan Ghosh and S. Umesh},
  title = {data2vec-aqc: Search for the right Teaching Assistant in the Teacher-Student training setup},
  year = {2022},
  eprint = {2211.01246},
  archivePrefix = {arXiv},
  primaryClass = {cs.LG},
  url = {https://arxiv.org/abs/2211.01246}
}

@misc{baevski2022data2vec,
  author = {Alexei Baevski and Wei-Ning Hsu and Qiantong Xu and Arun Babu and Jiatao Gu and Michael Auli},
  title = {data2vec: A General Framework for Self-supervised Learning in Speech, Vision and Language},
  year = {2022},
  eprint = {2202.03555},
  archivePrefix = {arXiv},
  primaryClass = {cs.LG},
  url = {https://arxiv.org/abs/2202.03555}
}

@misc{IITM_IndicTTS,
  author = {{Speech Technology Consortium, IIT Madras}},
  title = {IndicTTS: A Text-to-Speech Dataset for Indian Languages},
  year = {2025},
  url = {https://www.iitm.ac.in/donlab/indictts/database},
  note = {Accessed: 2025-09-29}
}

@misc{indic_conformer,
  author = {Tahir Javed and Kaushal Bhogale},
  title = {IndicConformer},
  howpublished = {\url{https://ai4bharat.iitm.ac.in/areas/model/ASR/IndicConformer}},
  note = {Accessed: 2025-07-22},
  year = {2025}
}

@misc{vaani-whisper,
  author = {{ARTPARK and IISc}},
  title = {Vaani Whisper Collection},
  howpublished = {\url{https://huggingface.co/collections/ARTPARK-IISc/vaani-whisper-67d4438f63799fb4eb94aec2}},
  note = {Accessed: 2025-07-22},
  year = {2025}
}

@misc{mucs,
  author = {{MUCS}},
  title = {MUCS ASR CHallenge},
  howpublished = {\url{https://navana-tech.github.io/MUCS2021/data.html}},
  note = {Accessed: 2025-07-22},
  year = {2025}
}

@misc{vaani2025,
  title={VAANI: Capturing the language landscape for an inclusive digital India}, 
  author={Sujith Pulikodan and Abhayjeet Singh and Agneedh Basu and Nihar Desai and Pavan Kumar J and Pranav D Bhat and Raghu Dharmaraju and Ritika Gupta and Sathvik Udupa and Saurabh Kumar and Sumit Sharma and Vaibhav Vishwakarma and Visruth Sanka and Dinesh Tewari and Harsh Dhand and Amrita Kamat and Sukhwinder Singh and Shikhar Vashishth and Partha Talukdar and Raj Acharya and Prasanta Kumar Ghosh},
  year={2026},
  eprint={2603.28714},
  archivePrefix={arXiv},
  primaryClass={eess.AS},
      url={https://arxiv.org/abs/2603.28714}, 
}

@standard{etsi_gsm610,
  title        = {Digital cellular telecommunications system (Phase 2); Full rate speech; Transcoding},
  organization = {ETSI},
  number       = {GSM 06.10},
  year         = {1991}
}

@standard{itu_g712,
  title        = {Transmission performance characteristics of pulse code modulation channels},
  organization = {ITU-T},
  number       = {G.712},
  year         = {2001}
}

@standard{itu_g722,
  title        = {7 kHz audio-coding within 64 kbit/s},
  organization = {ITU-T},
  number       = {G.722},
  year         = {2012}
}

@inproceedings{valin2012opus,
  author    = {Jean-Marc Valin and Koen Vos and Timothy Terriberry},
  title     = {Definition of the Opus Audio Codec},
  booktitle = {IETF RFC 6716},
  year      = {2012}
}

@book{smith2002digital,
  author    = {Julius O. Smith III},
  title     = {Digital Audio Resampling Home Page},
  year      = {2002},
  note      = {Stanford University, Online resource}
}

@book{oppenheim1999discrete,
  author    = {Alan V. Oppenheim and Ronald W. Schafer},
  title     = {Discrete-Time Signal Processing},
  publisher = {Prentice Hall},
  year      = {1999}
}

@inproceedings{liu2022voicefixer,
  author    = {Haohe Liu and Yi Yuan and Xubo Liu and Qiuqiang Kong and Mark D. Plumbley},
  title     = {VoiceFixer: Towards General Speech Restoration with Neural Vocoder},
  booktitle = {Proc. Interspeech},
  year      = {2022}
}

@inproceedings{liu2023audiosr,
  author    = {Haohe Liu and Xubo Liu and Qiuqiang Kong and Mark D. Plumbley},
  title     = {AudioSR: Versatile Audio Super-resolution at Scale},
  booktitle = {Proc. ICASSP},
  year      = {2023}
}

@inproceedings{pratap2020massively,
  title={Massively Multilingual ASR: 50 Languages, 1 Model, 1 Billion Parameters},
  author={Pratap, Vineel and others},
  booktitle={Interspeech},
  year={2020}
}

@article{li2014noise,
  title={An overview of noise-robust automatic speech recognition},
  author={Li, Jian and Deng, Li and Gong, Yu and Haeb-Umbach, Rudolf},
  journal={IEEE/ACM Transactions on Audio, Speech, and Language Processing},
  volume={22},
  number={4},
  pages={745--777},
  year={2014}
}

@inproceedings{graves2006ctc,
  author    = {Graves, Alex and Fern{\'a}ndez, Santiago and Gomez, Faustino and Schmidhuber, J{\"u}rgen},
  title     = {Connectionist Temporal Classification: Labelling Unsegmented Sequence Data with Recurrent Neural Networks},
  booktitle = {Proceedings of the 23rd International Conference on Machine Learning (ICML)},
  pages     = {369--376},
  year      = {2006},
  publisher = {ACM}
}

\end{document}